# Compressive mechanical response of graphene foams and their thermal resistance with copper interfaces

Wonjun Park,[1, 2, a)] Xiangyu Li,[3,2] Nirajan Mandal,[4] Xiulin Ruan,[3, 2] and Yong P. Chen[4, 2, 1, b)]

[1]*School of Electrical and Computer Engineering, Purdue University, West Lafayette, IN 47907, U.S.A.*
[2]*Birck Nanotechnology Center, Purdue University, West Lafayette, IN 47907, U.S.A.*
[3]*School of Mechanical Engineering, Purdue University, West Lafayette, IN 47907, U.S.A.*
[4]*Department of Physics and Astronomy, Purdue University, West Lafayette, IN 47907, U.S.A.*

**Abstract**

We report compressive mechanical response of graphene foams (GFs) and the thermal resistance ($R_{TIM}$) between copper (Cu) and GFs, where GFs were prepared by the chemical vapor deposition (CVD) method. We observe that Young's modulus ($E_{GF}$) and compressive strength ($\sigma_{GF}$) of GFs have a power law dependence on increasing density ($\rho_{GF}$) of GFs. The maximum efficiency of absorbed energy ($\eta_{max}$) for all GFs during the compression is larger than ~0.39. We also find that a GF with a higher $\rho_{GF}$ shows a larger $\eta_{max}$. In addition, we observe that the measured $R_{TIM}$ of Cu/GFs at room temperature with a contact pressure of 0.25 MP applied increases from ~50 to ~90 mm$^2$K/W when $\rho_{GF}$ increases from 4.7 to 31.9 mg/cm$^3$.

[a)] Electronic mail: park249@purdue.edu
[b)] Electronic mail: yongchen@purdue.edu

Three-dimensional (3D) carbon structures have attracted much attention for the past decades due to unique structural properties (e.g., lightweight, flexibility), and excellent electrical and thermal properties. In particular, after the pioneering experiments on graphene,[1] there are a lot of efforts to develop 3D graphene based structures by using chemically exfoliated graphene derivatives such as graphene oxide (GO) or reduced GO.[2] It has been demonstrated that this approach (assembling graphene derivatives to 3D structures) can be useful for some applications.[2] However, the quality of graphene derivatives is usually inferior to pristine graphene or graphite due to significant defects caused by the chemical exfoliation process. In addition, the 3D structures prepared by assembling graphene derivatives inevitably include many thermal or electrical interfaces (between graphene and graphene, or graphene and polymer matrices in their composites), causing deterioration of electrical and thermal transport in freestanding structures or their polymer composites.

On the other hand, 3D interconnected graphene structures grown by the chemical vapor method (CVD) such as graphene (or graphitic) foams (GFs), in which the quality of graphene is comparable to pristine graphene or graphite, have been recently developed[3] and enabled us to achieve continuous structures with less redundant interfaces, providing good electrical and thermal conduction paths in ultra-lightweight structures.[3,4] GFs grown by CVD have been demonstrated to be useful for applications for elastic polymeric conductors, electromagnetic interference (EMI) shielding, gas sensors, electrochemical applications (e.g., biosensors, electrodes for batteries and supercapacitors), packaging, and thermal interface materials (TIMs).[5–8]

In particular, 3D GFs or their composites are promising materials for TIM applications and packaging applications due to their excellent thermal and mechanical properties. One can achieve efficient heat conduction across thermal interfaces by filling in surface irregularities (e.g., air gaps) at the interfaces with TIMs. In general, many TIMs are composite materials consisting of fillers (e.g., metal, ceramic, or carbon based particles) and matrices (e.g., hydrocarbon oil, silicone oil, epoxy, or phase change materials). The thermal resistance ($R_{TIM}$) of those composite TIMs at interfaces can be 2-60 mm$^2$K/W.[9, 10] On the other hand, the R$_{TIM}$ of a graphite paper or a carbon nanotube paper can be 20-170 mm$^2$K/W, which is usually higher than that of those composite types of TIMs by one order of magnitude.[11–13] In this work, we studied the compressive mechanical response of GFs with various densities, and the R$_{TIM}$ between copper (Cu) and compressed GFs, where GFs were prepared by the CVD method. Some studies on GFs prepared by the CVD method have been reported but most of these studies have focused on GFs with a limited range of densities. Density-dependent mechanical response and R$_{TIM}$ for CVD-grown GFs (at the density range of a few mg/cm$^3$ to tens of mg/cm$^3$) have not been reported, despite the promise of thermal interface applications and many mechanical applications. Such studies as reported in this work will provide insights about the mechanical response of GFs under compression and the R$_{TIM}$ for particular packaging applications or thermal interface applications.

We prepared GFs using a similar method as described in Ref.3. Nickel (Ni) foams (70 pores per inch and 3 mm-thick) were cleaned with 20% acetic acid solution for 1 hour and rinsed with deionized water. Pre-cleaned Ni foams were loaded into a horizontal CVD furnace chamber and annealed at 1050 °C for 1 hour with a flow of H$_2$ (30 sccm) and Ar (170 sccm). GF was grown on the Ni foam with a flow of CH$_4$ (20 sccm), H$_2$ (20 sccm), and Ar (210 sccm) for 8-120 minutes. After the growth, we coated the GF/Ni foam with polymethyl methacrylate (PMMA) to protect the GF from an etching process. The Ni foam was etched by 1 M (molar concentration) iron nitrate nonahydrate (>98%, Sigma Aldrich) solution at 90 °C for 2 days, and additionally etched by 1 M hydrochloric acid solution at 90 °C for 2 days. After the etching process, we removed PMMA using acetone.

Microscopic structures of GFs were studied using a scanning electron microscope (SEM) (FEI Nova 200 NanoLab DualBeam SEM). X-ray diffraction spectra of GFs were collected using a X-ray diffractometer (Bruker D8 Focus) with a Cu Kα radiation source (wavelength~1.54 Å). We performed Raman spectroscopy on GFs using a confocal Raman microscope (Horiba Jobin Yvon Xplora) with a 100× objective lens and a 532 nm laser (~1 mW of incident power on the sample). We conducted the compression test using a universal mechanical test system (Instron 3345) with a 50 N load cell. We prepared a square-shaped GF (10 mm × 10 mm × 3 mm) and placed the GF between two parallel metal plates (each metal plate is mechanically attached to upper and lower anvil platens, respectively) as shown in the inset in Fig. 2(a). Before the compression test started, we performed calibration and zeroing of the system. The compression test was performed at a strain rate of 3 μm sec$^{-1}$. The data for displacement and force was recorded using the Bluehill software (provided by Instron). On average, four samples for each density were studied to extract the Young's modulus ($E_{GF}$) and compressive strength ($\sigma_{GF}$) of GFs. We measured the $R_{TIM}$ between Cu and compressed GFs while changing the contact pressure from 0.14 to 1 MPa, using a modified ASTM 5470 method described in Ref.14. In short, GFs were placed between upper and lower heat flux meters (Cu rods with a diameter of 19.05 mm), where the each surface of the Cu rods was polished using 1μm alumina particles (root mean square roughness of the Cu surface~300 nm). A heating power was supplied by a 50 ohm cartridge heater attached on the upper heat flux meter, and a cooling power was applied by flowing water with a controlled fluidic pressure. Temperatures along the heat flux direction were measured using six Pt resistance temperature detectors (RTDs) when a steady state condition is reached. The heat flux and temperature were controlled by a temperature controller (Lake Shore 340).

Figs.1(a) and 1(b) show microscopic structures of GFs with different densities ($\rho_{GF}$), confirming 3D interconnected structures (consisting of interconnected graphitic ligaments) with a pore size of 100-500 μm. In general, we find that the surface morphology of low-density GFs is smoother than that of high-density GFs, as also observed in the previous studies.[4] We studied the crystal structure of GFs using XRD as shown in Fig. 1(c). We observe two significant peaks at 2θ=26.5° and 54.6°, originating from (002) and (004) planes in graphite, respectively (JCPDS 75-1621).[15] The interplanar spacing is calculated to be ~3.36 Å, similar to that of highly oriented pyrolytic graphite (HOPG).[4] In addition, the representative Raman spectrum of a GF in Fig. 1(d) demonstrates a low defect density, as shown in the negligible intensity of the Raman "D" peak, indicating a high-quality graphitic structure.

It is important to understand the mechanical response of GFs for some applications such as packaging or TIMs. We studied the stress ($\sigma$)-strain ($\varepsilon$) behavior of GFs as shown in Fig. 2(a), by applying the compressive load. The compressive mechanical response for GFs exhibits three distinct regions (labeled in Fig. 2(a)): a linear elastic region, a plateau region (plastic deformation region), and a densification region. We observe that the linear elastic region occurs below $\varepsilon\sim0.1$ for GFs with various $\rho_{GF}$, which is related to elastic bending of graphitic ligaments in GFs. Unlike the elastic region, the plateau region, where $\sigma$ is almost independent of $\varepsilon$, originates from plastic collapses of graphitic ligaments. We also find noticeable peaks for high-density (e.g., 31.5 mg/cm$^3$) GFs in the plateau regime, indicating multi-stage collapses of those graphitic ligaments, similar to the previous report.[16] Above the plateau region, it is observed that $\sigma$ sharply increases with increasing $\varepsilon$, suggesting that graphitic walls touch together and start to be packed. Thus, a higher $\sigma$ is required to displace GFs, resulting in decreasing efficiency for the GFs to absorb energy during the further compression, as shown below.

The compressive Young's modulus ($E_{GF}$, the slope of $\sigma$-$\varepsilon$ in the elastic regime) of GFs increases from ~16 kPa to ~418 kPa with increasing $\rho_{GF}$ of GFs from ~4 to ~32 mg/cm$^3$, as shown in Fig. 2(b). The trend shows a power law dependence on $\rho_{GF}$ (Fig. 2(b) inset). Similarly, the compressive strength ($\sigma_{GF}$, the stress at which the plastic deformation starts) of GFs as a function of $\rho_{GF}$ also exhibits a power law dependence on $\rho_{GF}$ as shown in Fig. 2(c). The $\rho_{GF}$-dependent $E_{GF}$ (or $\sigma_{GF}$) can be explained by the Gibson-Ashby (GA) model.[17–19] For our GFs, which are similar to plastic open-cell structures, $E_{GF}$ and $\sigma_{GF}$ follow $(E_{GF}/E_s)=C_1(\rho_{GF}/\rho_s)^n$ and $(\sigma_{GF}/\sigma_s)=C_2(\rho_{GF}/\rho_s)^m$, respectively, where $\rho_s$, $E_s$, and $\sigma_s$ are the density (~2.2 g/cm$^3$), Young's modulus (~20 GPa),[20, 21] and compressive strength of CVD-grown bulk graphite, respectively, and $C_1$ and $C_2$ are constants. The fitting of data yields n~1.6 and m~1.3 (as shown in the insets in Figs. 2(b) and 2(c)), respectively, compared to the expected n=2 and m=1.5 for the theoretical GA model.[18] Power law dependence of the Young's modulus has been found in various carbon based 3D structures but the experimentally measured exponent n could vary from ~1 to ~5, indicating the exponent strongly depends on structural details.[22]

The energy absorption properties of GFs with various $\rho_{GF}$ could provide fundamental information for designing proper cushion or packaging materials. Fig. 3(a) shows representative energy absorption diagrams (W/$E_s$ vs. $\sigma/E_s$) for GFs with various $\rho_{GF}$, where W(=$\int_0^{\varepsilon_m} \sigma(\varepsilon)d\varepsilon$) is the energy absorption capacity (absorbed energy per unit volume) up to a strain $\varepsilon_m$ (which is varied from 0 to 0.9 to calculate W($\varepsilon_m$) from the data in Fig. 2(a), and correspondingly $\sigma=\sigma(\varepsilon_m)$).[18] For GFs with different $\rho_{GF}$, we observe that W/$E_s$ gradually increases with increasing $\sigma/E_s$, and then the sharp increase (particularly pronounced in high-density GFs) of W/$E_s$ occurs in the plateau region (the plastic deformation region in $\sigma$-$\varepsilon$ curves). Above the sharp increase, the saturation region of W/$E_s$ is observed, implying the absorbed energy does not change significantly above the onset of densification. In general, to design cushion materials for packaging, it is believed that the optimal energy absorption can be achieved around the onset of the densification at a given density. In particular, around the onset of densification, W/$E_s$ of a GF with $\rho_{GF}$=31.5mg/cm$^3$ is larger by one order of magnitude than that of a GF with $\rho_{GF}$=4 mg/cm$^3$, implying a larger energy can be absorbed in GFs with a higher $\rho_{GF}$. It is observed that the densification for all GFs occurs around $\varepsilon_m$=0.6-0.7.

However, in order to make fair comparisons of the energy absorption properties for GFs with various $\rho_{GF}$, the efficiency of absorbed energy during compression should be considered. Figs. 3(b) and 3(c) show the energy absorption efficiency ($\eta$=W/$\sigma$, the ratio of the energy absorbed by GFs per unit volume to $\sigma$ applied) and the ideality factor ($\beta$=W/($\sigma\varepsilon_m$), the ratio of the energy absorbed by GFs up to $\varepsilon_m$ with $\sigma=\sigma(\varepsilon_m)$ to that by an ideal plastic foam with constant $\sigma$), respectively.[23] It is also found that $\eta$ and $\beta$ dramatically increase with increasing $\sigma$ until GFs enter the densification regime, and then more gradually decrease beyond this regime, showing our GFs could be more suitable for applications requiring a low stress below ~20 kPa. As shown in Fig. 3(d), for our GFs, the average values (among multiple samples tested) for $\eta_{max}$ and $\beta_{max}$ are larger than 0.39 and 0.68, respectively, which are comparable to or higher than those of many polymeric foams.[23] The highest $\eta_{max}$ (~0.5) and $\beta_{max}$ (~0.98) are found at GFs with the highest $\rho_{GF}$ (~32 mg/cm$^3$), implying the compression response of GFs with a higher $\rho_{GF}$ becomes close to an ideal plastic foam ($\beta=1$).

In order to evaluate usefulness of GFs as thermal interface materials, we investigate thermal resistance at interfaces ($R_{TIM}=2R_{int}+t/\kappa$, where $R_{int}$ is the interfacial thermal resistance between Cu and compressed GFs, t is the thickness of the compressed GFs, $\kappa$ is the thermal conductivity of the compressed GFs) between Cu and GFs with various applied contact pressures. The detailed measurement setup and process can be found in Ref.14. The measured $R_{TIM}$ of GFs as a function of $\rho_{GF}$ at room temperature with a contact pressure of 0.25 MPa is shown in Fig. 4(a) and the inset shows a representative temperature profile along the upper heat flux meter (Cu)/GF($\rho_{GF}$=4.7 mg/cm$^3$)/lower heat flux meter (Cu). We observe that $R_{TIM}$ of GFs increases from ~50 to ~90 mm$^2$K/W when $\rho_{GF}$ increases from 4.7 to 31.9 mg/cm$^3$. $R_{TIM}$ of our GF with 4.7 mg/cm$^3$ is larger by at least a factor of 7~10 than that of GFs measured at 0.23 MPa in a previous report.[8] We speculate that the portion of bulk contribution (t/$\kappa$) is more significant than that in Ref.8, possibly due to our use of GFs with a higher $\rho_{GF}$ giving a larger t at a given contact pressure (but a density information was not given in Ref.8), or use of thicker pre-compressed GFs (3mm-thick). We also studied $R_{TIM}$ of GFs with various $\rho_{GF}$ as a function of the contact pressure applied, showing decrease in $R_{TIM}$ with increasing the pressure applied, related to decrease in thickness (t) of the compressed GFs and increase in the real contact area between Cu and the compressed GFs at a higher contact pressure. In general, low-density GFs have larger compressibility, resulting in a larger contact area and a smaller t at a given contact pressure.

In summary, we investigated the compressive mechanical response of GFs and thermal resistance ($R_{TIM}$) at interfaces between GFs and Cu in order to evaluate usefulness of GFs for packaging or thermal interface application. We observe that Young's modulus ($E_{GF}$) and compressive strength ($\sigma_{GF}$) of GFs follow the power dependence on the density ($\rho_{GF}$) of GFs with the exponent n=~1.6 and m=~1.3, respectively. In addition, we find that the maximum efficiency of absorbed energy ($\eta_{max}$) and the maximum ideality factor ($\beta_{max}$) of GFs during compression can be as high as $\eta_{max}$~0.5 and $\beta_{max}$~0.98, which are comparable to or higher than those of many polymeric foams. We also characterized the $R_{TIM}$ at interfaces between Cu and GFs with various $\rho_{GF}$. We measure $R_{TIM}$ of a GF with $\rho_{GF}$ =4.7 mg/cm$^3$ to be ~50 mm$^2$K/W with 0.25 MPa applied and it can be ~25 mm$^2$K/W with 1MPa applied, which is comparable to commercial graphite papers (20-170 mm$^2$K/W).[11, 12] The measured $R_{TIM}$ is found to be larger than that of GFs in the previous report, possibly due to the significant bulk contribution (t/κ). However, with a proper selection of templates for the CVD process, a proper selection of $\rho_{GF}$, or a use of polymers, we believe that $R_{TIM}$ can be reduced further and the energy absorption property can be tunable as well, possibly making GFs a promising material for many applications including packaging and thermal interface materials.

**Acknowledgement**

We thank Dr. Rebecca Kramer at Purdue University for access to Instron 3345 and also thank Ms. Michelle Yuen at Purdue University for helping us with mechanical characterizations. This work was partly supported by the Purdue Cooling Technologies Research Center (CTRC, a National Science Foundation (NSF) industry/university cooperative research center), and NSF Civil, Mechanical and Manufacturing Innovation (CMMI) program (1538360).

**Figures and captions**

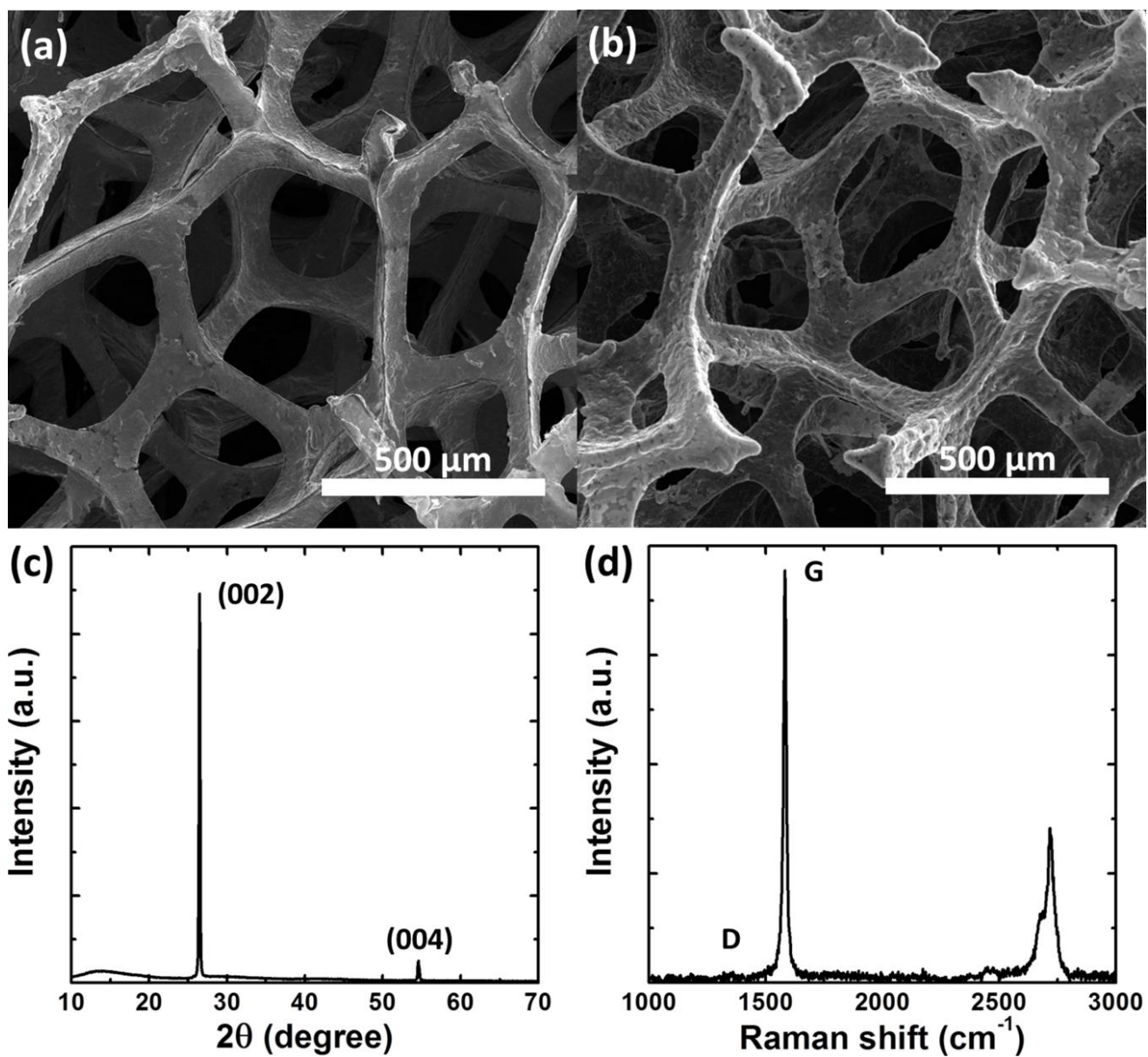

FIG. 1. (a, b) SEM images of as-grown graphene foams (GFs) with a density ($\rho_{GF}$) of (a) 4 and (b) 13.5 mg/cm$^3$. (c) Representative XRD spectrum of an as-grown GF. (d) Representative Raman spectrum of an as-grown GF.

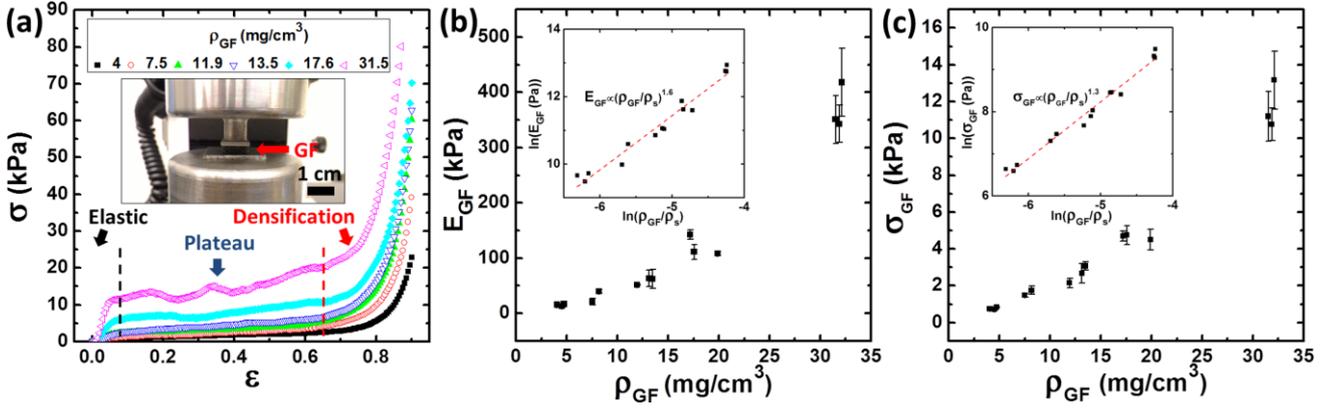

FIG. 2. (a) Compressive stress (σ)-strain (ε) curves of graphene foams (GFs) with various densities ($\rho_{GF}$). Inset: illustration of the compression test setup. (b) Compressive Young's modulus ($E_{GF}$) of GFs as a function of $\rho_{GF}$. Inset: $\ln(E_{GF})$ versus $\ln(\rho_{GF}/\rho_s)$, where $\rho_s$ is the density of bulk graphite. (c) Compressive strength ($\sigma_{GF}$) of GFs as a function of $\rho_{GF}$. Inset: $\ln(\sigma_{GF})$ versus $\ln(\rho_{GF}/\rho_s)$.

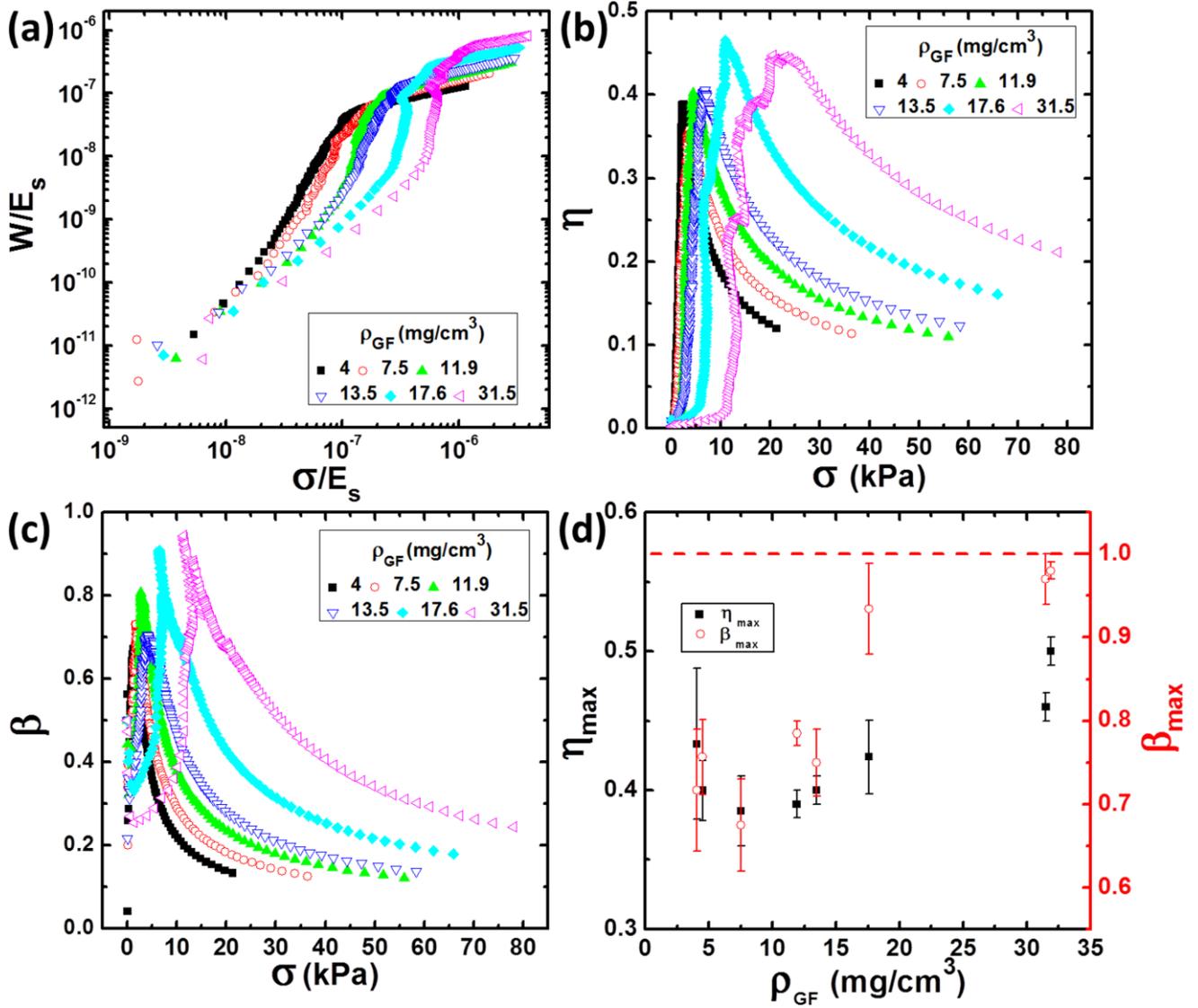

FIG. 3. (a) Energy absorption diagrams ($W/E_s$ versus $\sigma/E_s$) for graphene foams (GFs) with various densities ($\rho_{GF}$), where W is the absorbed energy per unit volume, $E_s$ is the Young's modulus of bulk graphite, and $\sigma$ is the stress. (b) Energy absorption efficiency ($\eta=W/\sigma$) of GFs with various $\rho_{GF}$ as a function of $\sigma$. (c) Ideality factor ($\beta=W/(\sigma\varepsilon_m)$, where $\varepsilon_m$ is varied from 0 to 0.9) of GFs with various $\rho_{GF}$ as a function of $\sigma$. (d) Maximum $\eta$ ($\eta_{max}$) and maximum $\beta$ ($\beta_{max}$) of GFs as a function of $\rho_{GF}$. The red dashed line represents an ideal plastic foam ($\beta=1$).

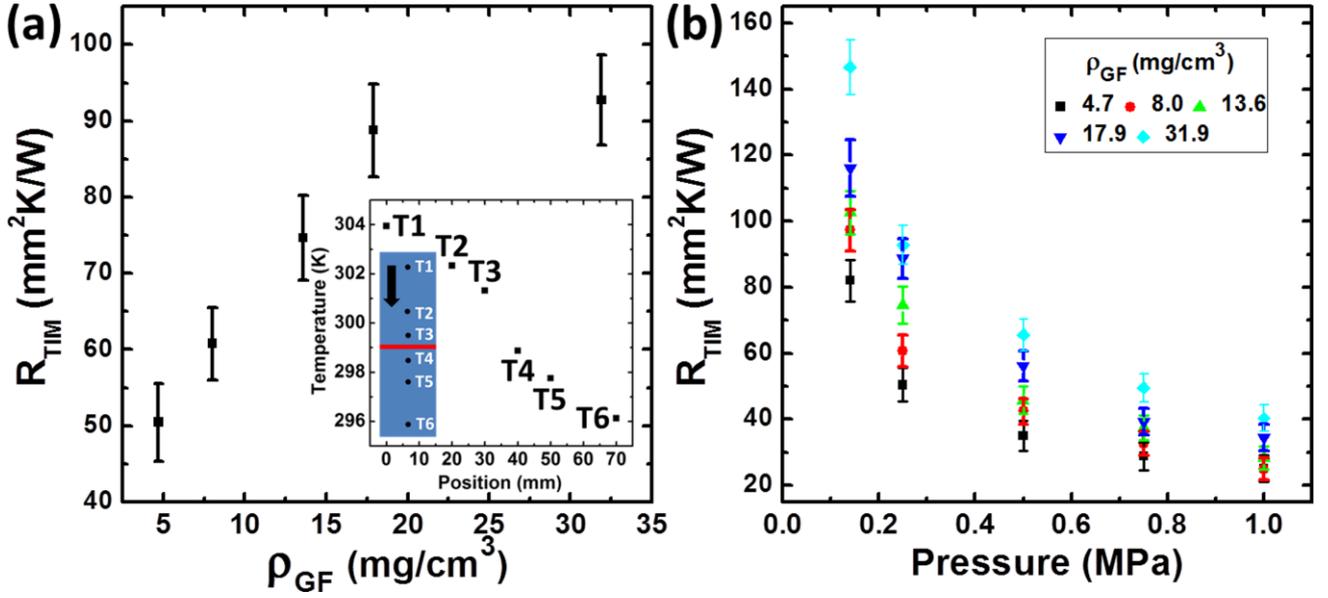

FIG. 4. (a) Thermal resistance ($R_{TIM}$) between graphene foams (GFs) and copper (Cu) as a function of density ($\rho_{GF}$) with a pressure of 0.25 MPa applied at 300 K. Inset figure shows a representative temperature profile along the upper heat flux meter (Cu)/GF($\rho_{GF}$=4.7 mg/cm$^3$)/lower heat flux meter (Cu), and the temperatures (from T1 to T6) are read by RTDs in the heat flux meters (blue colored), where the black arrow represents the heat flux direction and the red line represents the compressed GF (the thickness (t) of the compressed GFs is usually less than 500 μm). (b) $R_{TIM}$ of GFs with various $\rho_{GF}$ as a function of pressure applied at 300 K.